\documentstyle[epsfig,12pt]{article}

\newcommand{\bce}{\begin{center}}
\newcommand{\ece}{\end{center}}
\newcommand{\beq}{\begin{equation}}
\newcommand{\eeq}{\end{equation}}
\newcommand{\bea}{\vspace{0.25cm}\begin{eqnarray}}
\newcommand{\eea}{\end{eqnarray}}

\newcommand{\bk}{{\bf k}}

\newcommand{\br}{{\bf r}}

\newcommand{\ba}{\begin{array}}
\newcommand{\ea}{\end{array}}

\newcommand{\ket}[1]{| {#1} \rangle}

\newcommand{\ave}[1]{\langle {#1} \rangle}

\newcommand{\doublespace}{
\renewcommand{\baselinestretch}{1.6}\large\normalsize}

\def\lsim{\mathrel{\rlap{\lower4pt\hbox{\hskip1pt$\sim$}}
\raise1pt\hbox{$<$}}}         
\def\gsim{\mathrel{\rlap{\lower4pt\hbox{\hskip1pt$\sim$}}
\raise1pt\hbox{$>$}}}         

\def\beq{\begin{equation}}
\def\endeq{\end{equation}}
\def\arr{\begin{eqnarray}}
\def\endarr{\end{eqnarray}}
\makeindex
\advance\oddsidemargin  by -1in
\advance\evensidemargin by -1in
\advance\topmargin      by -0.5in
\begin{document}

\vspace{2.0cm}

\vspace{1.0cm}

\begin{center}
{\Large \bf
Color dipoles, PCAC and  Adler's theorem}

\vspace{1.0cm}

{\large\bf R.~Fiore$^{1 \dagger}$ and V.R.~Zoller$^{2 \ddagger}$}

\vspace{1.0cm}
$^1${\it Dipartimento di Fisica,
Universit\`a     della Calabria\\
and\\
Istituto Nazionale
di Fisica Nucleare, Gruppo collegato di Cosenza,\\
I-87036 Rende, Cosenza, Italy}\\
$^2${\it
ITEP, Moscow 117218, Russia\\}
\vspace{1.0cm}
{ \bf Abstract }\\
\end{center}
Being reformulated in the color dipole basis of  small-$x$ QCD
Adler's theorem establishes a connection  between perturbative
and non-perturbative descriptions of DIS  and quantifies the effect
of non-perturbative dynamics on would-be-perturbative observables.
 In particular, it provides
 a quantitative measure  of the non-perturbative  
influence on  the longitudinal 
structure function in  charged current DIS and
imposes stringent constraints on non-perturbative parameters of 
color dipole models. Our analysis  calls for new experimental tests of 
Adler's theorem in diffractive
neutrino scattering.  
\doublespace
\vskip 0.5cm \vfill $\begin{array}{ll}
^{\dagger}\mbox{{\it email address:}} & \mbox{fiore@cs.infn.it} \\
^{\ddagger}\mbox{{\it email address:}} & \mbox{zoller@itep.ru} \\
\end{array}$
\pagebreak

\section{Introduction}
Adler's theorem  \cite{Adler} connects charged current DIS
(deep inelastic scattering) with
soft hadronic physics. Here we study its efficiency as a constraint on
parameters of color dipole models intended to describe phenomenologically both
soft and hard dynamics of DIS processes.  We
focus on the vacuum exchange dominated
leading $\log(1/x)$ region of large  Regge parameter
$x^{-1}\gsim 10^2$ which is
\beq
x^{-1}= {2m_N\nu\over m_A^2+Q^2}.
\label{eq:Regge}
\eeq
In (\ref{eq:Regge}) $\nu$ and $Q^2$ are the laboratory frame energy and
 virtuality of the  probe, respectively. The parameter  $m_A\simeq 1$ GeV
serves to define
the mass scale in the light-flavor axial channel. 
We base our consideration on
the color dipole (CD) approach to the  BFKL \cite{BFKL},
evolution of small-$x$ DIS \cite{NZZBFKL,NZ94}.
In this approach the  interaction of high-energy  neutrino
with the target nucleon, viewed in the laboratory frame,
 derives  from the coherent
interaction of $q\bar q, q\bar q g,...$
states in the light-cone electro-weak (EW) boson.
At small $x$  the
color dipole size,  $\br$, of the constituent quark-antiquark pair
is conserved in the interaction process and
the absorption cross section  for the EW boson in the
helicity state $\lambda$ is
calculated as  the quantum mechanical expectation value of
the flavor independent CD cross section $\sigma(x,r)$,
\beq
\sigma_{\lambda}(x,Q^2)=
\ave{\Psi_{\lambda}(z,\br)|\sigma(x,r)|\Psi_{\lambda}(z,\br)}.
\label{eq:CDF}
\eeq
 The CD structure of the
 W boson  is described by the
 light-cone wave function (LCWF) of the  quark-antiquark state
$\Psi_{\lambda}(z,\br)$ \cite{FZ1,Kolya92}.
An interesting possibility to gain deeper insight into the
 dynamics
of small and large color dipoles is offered by the neutrino DIS in
the axial channel.
In the axial channel at $Q^2\to 0$ the light-cone wave function of the
longitudinal W boson is proportional of the divergence of the
axial-vector current,
$\Psi_L\propto \partial_{\mu}A_{\mu}$.
The PCAC (partially  conserved   axial-vector  current) relation \cite{Nambu},
$\partial_{\mu}A_{\mu}=m^2_{\pi}f_{\pi}\varphi$,
connects  via Adler's theorem \cite{Adler}
the longitudinal cross section $\sigma_L$ defined by Eq.(\ref{eq:CDF})
and  the on-shell pion-nucleon total cross section
$\sigma_{\pi}=\ave{\Psi{_\pi}(z,\br)|\sigma(x,r)|\Psi{_\pi}(z,\br)}$,
\beq
\lim_{Q^2\to 0}Q^2\sigma_{L}(x,Q^2)=g^2f^2_{\pi}\sigma_{\pi}(\nu).
\label{eq:PS}
\eeq
In Eq.(\ref{eq:PS}) $\nu$ and $x$ are linked by the  Eq.(\ref{eq:Regge}).
The weak charge $g$ in (\ref{eq:PS}) is
related to the Fermi coupling constant $G_F$,
\beq
{G_F\over \sqrt{2}}={g^2\over m^2_{W}},
\label{eq:GF}
\eeq
and  $f_{\pi}\simeq 131$ MeV is the pion decay
constant.
The Eq.(\ref{eq:PS}) connects  absorption  cross sections of two high-energy
projectiles having  very different CD structure, the pointlike $W$ and
the non-pointlike $\pi$. While  the light-cone wave function
of the
 pion $\Psi_\pi(z,\br)$
is smooth and finite at small $r$
\footnote {for a review of the dominance of the soft LCWF and references
see \cite{Diehl,Radyushkin}},
the EW boson wave function is singular,
$\Psi_L\sim r^{-1}$. 
This singularity is  a legitimate pQCD effect and
it  makes evaluation of hard contributions to
$Q^2\sigma_{L}(x,Q^2)|_{Q^2\to 0}$ more reliable.
In particular, this singularity uniquely predicts that the small dipole (hard)
contribution to $\sigma_L$ is
much stronger than to $\sigma_{\pi}$.

Since the pioneering paper
\cite{PS}, where the  axial-vector meson dominance (AVMD) was  suggested,
Adler's theorem has been considered as a relation between the
higher mass contributions to the axial current and the pion-nucleon
 cross section.
 The  AVMD  was successfully
applied to the analysis of the  coupled-channel problem of
 neutrino-nucleus interactions in
Refs. \cite{BelKop,Kop,KopMar} where  the  mechanism of the   $\rho\pi$
dominance was analyzed in detail.  In Ref. \cite{LonThom}
the PCAC component of the cross section was identified with the
longitudinal part of the $a_1$-meson and  the constraint for the
longitudinal AVMD DIS structure function was obtained.
Thus, in the AVMD representation Adler's theorem relates so to say
soft physics to soft physics, the pion to higher axial-vector excitations.
Only in the CD basis Adler's theorem gains its rightful  heuristic power.
It establishes a connection  between perturbative
and non-perturbative processes and shows the effect, if not reveals a 
mechanism,
of non-perturbative dynamics on would-be-perturbative observables.
 In particular, it provides
 a quantitative measure  of the non-perturbative  
influence on  the longitudinal 
structure function of the light-flavor charged current (CC) DIS.
Below we discuss the origin and consequences of this observation.

\section{Adler's theorem and the axial  mass scale}
The color dipole (CD) approach
\cite{NZ91,M} proved to be
very successful in describing  of inclusive and diffractive
electroproduction DIS data in the vector channel down to $Q^2\sim m_q^2$,
where $m_q$ is the constituent quark mass
(for the review see \cite{HEBECKER} and also \cite{GBW}).
The mass scale in the vector channel
is fixed by the mass of lightest vector mesons, $m_V\sim 1$ GeV.
In the axial channel the spectrum of hadronic excitations starts with
the nearly massless pion. To get an idea of the characteristic  axial
 mass scale let us turn to Adler's theorem.
Following Adler \cite{Adler}, consider the particular case of
forward lepton  production in the reaction of neutrino-nucleon scattering
\beq
\nu(k)+N(p)\to l(k^{\prime})+X(p_X)
\label{eq:NU-N}
\eeq
in  the limit $Q^2=-q^2\to 0$ and 
 suppose that  ${k^{\prime}}^2=m^2_l=0$.
  In (\ref{eq:NU-N}) $p_X$
is the 4-momentum of the final hadronic state $X$,
$k$ and $k^{\prime}$ are the
4-momenta
of the neutrino  and final lepton and $q=k-k^{\prime}$ is the 4-momentum
carried by W-boson.
The amplitude of  the process  (\ref{eq:NU-N})
is
\beq
M= {G_F\over \sqrt{2}}l_{\mu}\ave{X|J_\mu  |p},
\label{eq:M}
\eeq
The  massless leptonic current $l_{\mu}$ is conserved and at $Q^2\to 0$ is
proportional to $q_{\mu}$,
\beq
l_{\mu}=\bar u_l(k^{\prime})\gamma_\mu(1-\gamma_5)u_{\nu}(k)=
{4\sqrt{1-y}\over y}q_{\mu}
\label{eq:Lepton}
\eeq
where $y=pq/kp=\nu/E$. The divergence
of the vector component of the  hadronic current
$J_{\mu}=V_{\mu}-A_{\mu}$
 is supposed to be zero.
Let $\ave{X|A_\mu|p}=M_{\mu}$ be
the sum of the pion pole term and
 amplitudes of higher axial-vector hadronic states, $\ket{a}$,
\beq
M_{\mu}=M^{\pi}_{\mu}+\sum_{a\neq \pi} M^a_{\mu}.
\label{eq:SIGR1}
\eeq
In  (\ref{eq:SIGR1})
\beq
M^{\pi}_{\mu}=iq_\mu f_\pi{{1 }\over {Q^2+m^2_\pi}}M_\pi\,.
\label{eq:SIGR2}
\eeq
Since $q_\mu l_\mu=0$, the pion pole does not contribute
to $l_{\mu}M_{\mu}$ and $M$ is saturated by higher
axial-vector states \cite{Bell}
\beq
M={G_F\over\sqrt{2}}\sum_{a\neq \pi}l_{\mu}M^a_{\mu}.
\label{eq:SIGR3}
\eeq
So, the mass scale in the axial channel has been determined, in fact,
in experiments on the diffraction dissociation of high-energy  pions,
 where the
mass spectrum of final
$a_1,\rho\pi, \pi\pi\pi...$ states was measured
(see \cite{Daum} and  also the discussion  in \cite{BelKop}).

Adler's amplitude is linear in the divergence of the axial current. 
To this accuracy, 
making use of $q_{\mu}M_{\mu}=0$ and
\beq
\sum_{a\neq \pi}q_{\mu}M^a_{\mu}=if_\pi M_\pi
\label{eq:GT}
\eeq
- the Goldberger-Treiman conspiracy \cite{GT}, - yields 
the neutrino amplitude which  is multiple of the pion amplitude
\beq
M ={iG_Ff_{\pi}\over \sqrt{2}}{4\sqrt{1-y}\over y} M_\pi.
\label{eq:QA}
\eeq
Summing over all final hadronic states
one arrives  at
Adler's  relation between $\sigma_{\pi}(\nu)$ and the
double differential cross section of the process (\ref{eq:NU-N}),
\bea
{d\sigma\over dydQ^2}\Big|_{Q^2=0}={\pi\over (pk)}
{1\over(4\pi)^3}\sum_X|M|^2(2\pi)^4\delta^{(4)}(p+q-p_X)
\nonumber\\
={G^2_F\over 2\pi^2}{1-y\over y}
f^2_\pi\sigma_\pi(\nu)
\label{eq:DSDYDQ2}
\eea

\section{CD models  and the axial mass scale again}
CD models rely upon the small-$x$ flux-cross section factorization,
\beq
yQ^2{d\sigma\over dydQ^2}=f_\lambda\sigma_\lambda
\label{eq:FLUXSEC}
\eeq
The  fluxes $f_\lambda$ and cross sections $\sigma_\lambda$
 depend on the polarization state
$\lambda$ of the EW boson.
In  the $W$-proton collision frame $q_\mu=(\nu,0,0,q_z)$.
and the 4-vector of the so called  longitudinal  polarization,
which we are interested in, is
\beq
s_\mu={1\over \sqrt{Q^2}}(q_z,0,0,\nu)={\sqrt{Q^2}\over \sqrt{(pq)^2+m_N^2Q^2}}
\left(p_\mu+{pq\over Q^2}q_\mu\right) .
\label{eq:SMU}
\eeq
Throughout this paper we use, following tradition,  the name ``longitudinal''
for the time-like vector $s_\mu$ and provide corresponding variables with
the label $L$.
Then, in close similarity with the QED flux of longitudinal photons,
the flux of $W_L$ bosons is
\beq
f_L={4\alpha_W\over\pi}{Q^4\over m^4_W}(1-y),
\label{eq:FLUXES}
\eeq
where $\alpha_W=g^2/4\pi$.
Applying the optical theorem to the amplitude
for Compton scattering of the  $W_L$  yields
\bea
{d\sigma\over dydQ^2}\Big|_{Q^2=0}
={G_F^2\over 2\pi^2}{(1-y)\over y}{Q^2\over g^2}
\sigma_L(x,Q^2)\Big|_{Q^2=0},
\label{eq:FLUXSEC1}
\eea
where, in the fixed-$\br$ representation,
\bea
\sigma_L(x,Q^2)=\ave{W|\br}\ave{\br|\hat\sigma|\br}
\ave{\br|W}\nonumber\\
=\int dz d^{2}{\bf{r}}
|\Psi_{L}(z,{\bf{r}})|^{2}
\sigma(x,r)\,.
\label{eq:FACTOR}
\eea
The Eq.(\ref{eq:PS} ) then follows from comparison of
Eqs. (\ref{eq:DSDYDQ2}) and (\ref{eq:FLUXSEC1}).
In Eq.~(\ref{eq:FACTOR}) $\ave{\br|\hat\sigma|\br}=\sigma(x,r)$,
$\hat\sigma$ is the  cross section operator and
$\ave{\br|W}=\Psi_{L}(z,{\bf{r}})$
 is the LCWF of
the $|u\bar d\rangle$ state with the $u$ quark
carrying fraction $z$ of the $W^+$ light-cone momentum and
$\bar d$ with momentum fraction $1-z$
(see \cite{FZ1} for details).
In \cite{FZ1} we used for $\Psi_L$
the notation $\Psi_0$.
The expansion (\ref{eq:SIGR1}) is written in the basis of physical hadrons
(mass operator
eigenstates) $\ket{a}$ related to fixed-${\br}$ states via
 $\ket{\br}=\sum_a\ave{a|\br}\ket{a}$. In this basis the diagonal matrix
elements of $\hat\sigma$ give the total cross section of $a-N$ scattering,
$\sigma_{a}=\ave{a|\hat\sigma|a}$ \cite{KolyaFacts}, and
  Eq.(\ref{eq:FACTOR}) turns  into
\bea
Q^2\sigma_L\Big|_{Q^2=0}=Q^2\ave{W|\br}\ave{\br|\hat\sigma|\br}
\ave{\br|W}\nonumber\\
=Q^2\sum_{a,a^{\prime}\neq \pi}\ave{W|a}\ave{a|\hat\sigma|a^{\prime}}
\ave{a^{\prime} |W}\nonumber\\
=Q^2\ave{W|\pi}\ave{\pi|\hat\sigma|\pi}\ave{\pi|W}
=g^2 f^2_\pi\sigma_\pi .
\label{eq:CDPCAC}
\eea
In (\ref{eq:CDPCAC}) the Goldberger-Treiman
conspiracy, Eq.(\ref{eq:GT}), was used to come from the second line 
 to the third one.

Notice that the sum over hadronic states $\ket{a}$ in (\ref{eq:CDPCAC})
does not include the pion, the
W-boson in the polarization state  $s_{\mu}$ (see Eq.(\ref{eq:SMU}))
 does not mix with the pion, $s_{\mu}q_{\mu}=0$.  Consequently,
the CD states described by   the light-cone
wave function $\Psi_L(z,\br)$ in Eq.(\ref{eq:FACTOR}) are  dual not to the
nearly massless  $\pi$-meson but
to ``normal''  axial-vector  hadronic  states ($a_1,\rho\pi,...$) of
a mass $\sim 1$ GeV. This observation justifies, in particular,
 the choice of the mass scale  $m_A=1$ GeV
in Eq.(\ref{eq:Regge}) and lends support to the CD
description  of  small-$x$ phenomena in the axial channel.

\section{$F_L$-$f_{\pi}$-$\sigma_{\pi}$-correlation in color dipole  basis.}

In terms of the longitudinal structure function
\beq
F_L(x,Q^2)={Q^2\over 4\pi^2\alpha_W}\sigma_L(x,Q^2),
\label{eq:FL}
\eeq 
Eq.(\ref{eq:CDPCAC}) can be rewritten as
\beq
F_L(x,0)= {f^2_\pi\over \pi}\sigma_\pi(\nu),
\label{eq:FLPCAC}
\eeq
where $\nu$ is related to $x$ by Eq.(\ref{eq:Regge}).
In (\ref{eq:FL}) $\sigma_L(x,Q^2)$ is defined by the CD factorization
equation (\ref{eq:FACTOR})  and  the light-cone density of CD  states  
$u\bar d, c\bar s,...$
is 
$$|\Psi_{L}(z,\br)|^2= |V_{L}(z,\br)|^2+ |A_{L}(z,\br)|^2$$
with
\bea
|V_L(z,{\bf r})|^2
={{2\alpha_W N_c}\over (2\pi)^2 Q^2}\left\{\left[2Q^2z(1-z)
+(m-\mu)[(1-z)m-z\mu]\right]^2K_0^2(\varepsilon r)
\right.
\nonumber\\
\left.
 +(m-\mu)^2
\varepsilon^2 K^2_1(\varepsilon r)\right\}
\label{eq:RHOS1}\\
|A_L(z,{\bf r})|^2
={{2\alpha_W N_c}\over (2\pi)^2 Q^2}\left\{\left[2Q^2z(1-z)
+(m+\mu)[(1-z)m+z\mu]\right]^2 K_0^2(\varepsilon r)
\right.
\nonumber\\
\left.
 +(m+\mu)^2
\varepsilon^2 K^2_1(\varepsilon r)\right\},
\label{eq:RHOS2}
\eea
where $m$ and $\mu$ are the quark and antiquark masses and 
$\varepsilon^2=z(1-z)Q^2+(1-z)m^2+z\mu^2$ \cite{FZ1,Kolya92}. 
At $Q^2\to 0$ and for equal  masses of  constituent quarks $m=\mu=m_q$,
the axial-vector light-cone  density of $u\bar d$ states does not depend
on $z$ and is as follows
\beq
|\Psi_{L}(r)|^2={{2\alpha_W N_c}\over \pi^2 }{m_q^2\over Q^2}
\left[m_q^2K_0^2(m_qr) + m_q^2K_1^2(m_qr)\right]
\label{eq:A2}
\eeq
At $y\lsim 1$, $K_0(y)\sim \log(1/y)$ and $K_1(y)\sim 1/y$. Then, from Eqs. 
(\ref{eq:FL}) and (\ref{eq:A2})
\beq
F_L(x,0)\sim {N_c m_q^2\over 2\pi^3}\int_0^{m_q^{-2}}
{dr^2\over r^2}\sigma(x,r).
\label{eq:FLESTIM}
\eeq
The CD cross section is 
$\sigma(x,r)=r^2C(x,r)$ with
$C(x,r)$ slowly varying with $r$. For small dipoles 
$C(x,r)$ depends on $r$ only logarithmically.
 For large dipoles, such that  $r > r_s$,
 $\sigma(x,r)$ saturates and  $C(x,r)=\sigma_s(x)/r^2$ \cite{NZ91}.
Therefore $F_L$ depends on several  non-perturbative parameters,
$m_q,r_s$ and $\sigma_s$ \footnote{There is, of course, one more 
(hidden) non-perturbative parameter -
the axial charge $g_A$.
The renormalization of $g_A$ is neglected here  and
the ratio $g_A/g_V$ for constituent quarks is assumed to be the same as
for current quarks, $g_A/g_V=1$.},
and, because
of the axial current non-conservation, it
is  sensitive to
the value of the constituent quark mass, 
\beq
F_L\propto m_q^2\sigma_s\log[1+1/(m_qr_s)^2].
\label{eq:FLPCAC1}
\eeq
 The sensitivity is lost, however,
for $m_q^2\gg r_s^{-2}$.

The {\sl rhs} of Eq.(\ref{eq:FLPCAC}) is known  experimentally
with high accuracy and,  in  view of all theoretical uncertainties with
non-perturbative effects, Eq.(\ref{eq:FLPCAC}) could  be considered as
a very useful constraint on parameters of CD models. Let us start, however,
with a self-consistency check and  before
imposing experimental bounds on $f_{\pi}$ and $\sigma_{\pi}$ let us 
evaluate $F_L$, $f_{\pi}$ and $\sigma_{\pi}$ 
within  the LCWF CD technique. Then, the deviation of the  ratio 
\beq
R_{PCAC}={{\pi}F_L(x,0)\over f^{2}_{\pi}\sigma_{\pi}(\nu)}.
\label{eq:RPCAC}
\eeq
from unity may serve as  the   measure of accuracy of the approach.
Because of (\ref{eq:FLPCAC1}) the dependence of $f_{\pi}$ and $\sigma_{\pi}$ 
on the constituent quark mass is of prime importance here.

Notice first,  that contrary to the  pointlike probe, the dipole size of
the non-pointlike  pion and, consequently, the value of 
 $\sigma_{\pi}$ is  determined not by $m_q$ but  by an 
additional  
parameter $R$ which introduces the  intrinsic  momentum cut-off.
 \footnote{ The radial part of $\Psi_{\pi}$ in  momentum space is 
$\Psi_{\pi}(M^2)\propto M^{-2}exp[-{1\over 8}R^2M^2]$ \cite{SNS} 
(see also \cite{Jaus}),
where $M$ stands for the invariant mass of the light-cone 
 $q\bar q$ state and
 $M^2={(m_q^2+\bk^2 )/z(1-z)}$.}  
The latter  removes the small-$r$ singularity
from $\Psi_{\pi}(z,r)$
and, simultaneously,  ensures the correct value of the  
charge radius of the pion. The dependence of $\sigma_{\pi}$ on $m_q$ 
appears to be  marginal and
\beq
\sigma_{\pi}=\ave{\Psi{_\pi}(z,\br)|\sigma(x,r)|\Psi{_\pi}(z,\br)}
\sim r^2_{\pi}C(x,r_{\pi}).
\label{eq:CSPI}
\eeq
The quantity which is very sensitive to $m_q$  
is the pion decay constant \cite{SNS,Jaus},
\beq
f_{\pi}={m_qN_c\over 4\pi^3}\int{dzd^2{\bk}\over z(1-z)}\Psi_{\pi}(M^2).
\label{eq:FPI}
\eeq
To a crude approximation, 
$f_{\pi}\propto m_q$ and  for 
$m_qR\ll 1$ (in \cite{SNS} $R=2.2$ GeV$^{-1}$) 
$$f_{\pi}\propto m_q\sqrt{\log(2/m_qR)}.$$ This gives 
a  chance  to
satisfy (\ref{eq:FLPCAC}) adjusting  properly non-perturbative  parameters. 
\begin{figure}[h]
\psfig{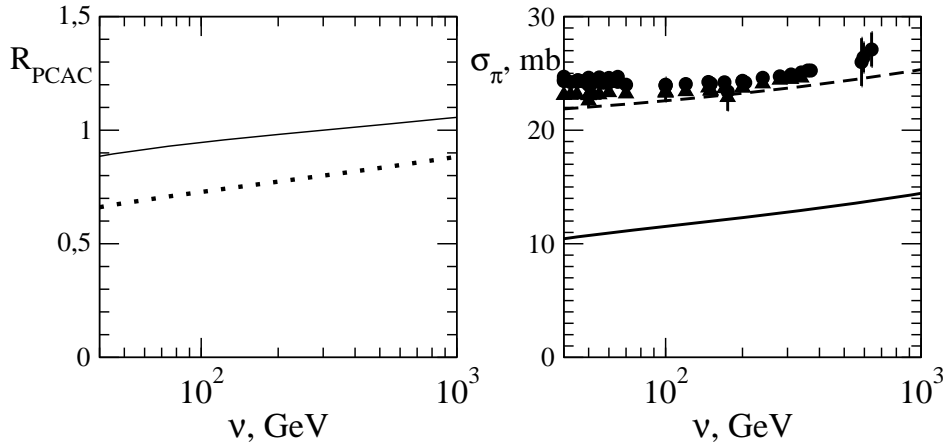}
\vspace{-0.5cm}
\caption{\small 
Left panel: the ratio $R_{PCAC}$ as a function of $\nu$ 
evaluated  with the pion LCWF 
of Ref.\cite{SNS}
 for $m_q=150$ MeV - thin solid line; 
the ratio $R_{PCAC}$ with the pion LCWF 
of Ref.\cite{Jaus} and
$m_q=250$ MeV - dotted line.  Right panel:
the CD BFKL evaluation of $\sigma_{\pi}(\nu)$ - dashed line.
Data points (triangles and circles) are measurements of  
total $\pi^+p$ and $\pi^-p$
cross sections, respectively \cite{PDG}.
Also shown by the
thick solid line is  the quantity $\sigma_{PCAC}(\nu)$ for the empirical value of $f_{\pi}$
and  with $F_L$ evaluated  
for $m_q=150$ MeV.}
\label{fig:adler}
\end{figure} 
 
Invoking the CD factorization,
which is valid for  soft as well as for hard
diffractive interactions,
 we evaluate the vacuum exchange contribution to $F_L(x,0)$
and $\sigma_{\pi}(\nu)$.   The structure function $F_L(x,0)$ comes from
Eqs.(\ref{eq:FACTOR},\ref{eq:FL},\ref{eq:A2}) with $m_q=150$ MeV, the value 
commonly used now in CD models
successfully tested against DIS data.
The $\log(1/x)$-evolution
of $\sigma(x,r)$ is
 described by the CD BFKL equation \cite{NZZPHL94,BFKLRegge}.
Corresponding boundary condition is found in \cite{Shad06}. With the pion LCWF of 
Ref.\cite{SNS}    we obtain $\sigma_{\pi}$ 
shown in  Fig.\ref{fig:adler} (right panel)  by the dashed line.
The $\nu$-dependence of $R_{PCAC}$ is  shown in Fig. \ref{fig:adler} (left panel)
by the thin solid line. With certain reservations about the slope of $R_{PCAC}(\nu)$ (see below)
we  conclude that our  CD model  successfully passed the consistency test. However,
this ``purely theoretical'' approach to Adler's theorem is not quite satisfactory.
The point is that the constituent quark with  $m_q=150$ MeV  amounts to $f_{\pi}=96$ MeV 
{\sl vs.} the empirical value $f_\pi=130.7\pm 0.46$ MeV  \cite{PDG}: not quite bad
for the model evaluation of the soft parameter $f_{\pi}$, although not satisfactory either. 
Within the model \cite{SNS} $f_{\pi}=131$ MeV corresponds  to $m_q=245$ MeV,
 the value 
which is very close to $m_q=250$ MeV of  Ref.\cite{Jaus}. In \cite{Jaus}   an oscillator type
ansatz for the pion LCWF was used and   good agreement of predictions of the model 
 with  both  the empirical
 value of  the pion decay constant and
the  charge radius of the pion was found.
  The ratio $R_{PCAC}$
evaluated with the pion LCWF of Ref.\cite{Jaus} is shown in Fig.\ref{fig:adler} (left panel)
by the dotted line.  
Evidently,  making the light quark heavier affects the distribution of 
color dipoles
in the light-cone $W_L$ boson in such a way that
the characteristic dipole sizes contributing to $F_L$, $r^2\lsim m_q^{-2}$ 
(mind also the small-$r$ singularity in $\Psi_L$),
becomes  much smaller than those contributing 
to $\sigma_{\pi}$, $r^2\sim r^2_{\pi}\simeq 1.2$ fm$^2$. 
The BFKL $\log(1/x)$-evolution  
of dipole cross sections  is characterized by
the exponent $\Delta(x,r)$  of the local $x$-dependence 
of $\sigma(x,r)$ 
\beq
\sigma(x,r)\propto \exp[-\Delta(x,r)\log(1/x)].
\label{eq:PREAS}
\eeq 
$\Delta(x,r)$ varies with $r$ and $\Delta(x,r_1)>\Delta(x,r_2)$ for $r_1<r_2$ 
\cite{NZZPHL94}. Hence, the  structure function $F_L(x,0)$ growing with
 $\nu$ faster than $\sigma_{\pi}(\nu)$ 
 in conflict with the requirement of Eq.(\ref{eq:FLPCAC}). 

 As we noted above Eq.(\ref{eq:FLPCAC}) can be considered as a condition on 
parameters of CD models. To see its restrictive power  in action
one can  construct the quantity  
\beq
\sigma_{PCAC}(\nu)=\pi F_L(x,0)/f^2_{\pi}
\label{eq:SPCAC}
\eeq
with the empirical  value of  $f_{\pi}=130.7$ MeV
and compare its $\nu$-dependence  with experimental data on 
$\sigma_{\pi}(\nu)$. 
Our $\sigma_{PCAC}(\nu)$ is  shown by the thick solid line in 
Fig.\ref{fig:adler} (right panel). It strongly undershoots 
$\sigma_{\pi}(\nu)$  thus indicating
 that $F_ L(x,0)$ evaluated with $m_q=150$ MeV fails to 
 satisfy the  Eq.(\ref{eq:FLPCAC}). 
 We tried also the  CD cross sections of Refs.\cite{GBW}. 
Corresponding  $F_L$ proved to be 
close to ours. 
Notice that, the discrepancy observed may have the same origin as 
the deficit of 
the differential cross section of diffractive vector meson production found 
in \cite{VecMesons}. That calls for 
better understanding of the infrared properties of the CD cross section.

From a different point of view, 
good agreement with data of both $\sigma_{\pi}(\nu)$ and $f_{\pi}$
spoiled, however, by the under-predicted $F_L(x.0)$ implies that
 non-perturbative interactions in the axial channel
blow-up the   dipole size   
 and, in the spirit of PCAC,
$\sqrt{Q^2}\Psi_L(z,\br)\to gf_{\pi}\Psi_{\pi}(z,\br)$ at $Q^2\to 0$.
\begin{figure}[h]
\psfig{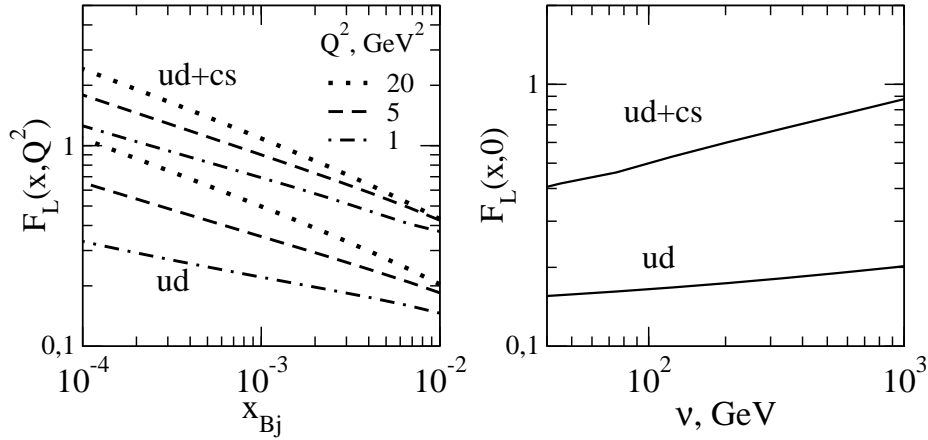}
\vspace{-0.5cm}
\caption{\small Left panel: three lower lines represent the $ud$-current
 contribution to $F_L(x,Q^2)$
as a function of Bjorken variable denoted by $x_{Bj}$ 
for different values of $Q^2$. 
Three upper lines correspond to the sum of $ud$- and $cs$-current  
contributions to  
$F_L$. Right panel: $F_L$ as a function of $\nu$ for $Q^2=0$.}
\label{fig:fl}
\end{figure}

\section{Weak current non-conservation 
and charm-strange dominance.}

One more concluding remark is on the role of charm-strange current.
The structure function $F_L(x,0)$
 shown in Fig.\ref{fig:adler} does not contain the 
$cs$-current contribution. 
The latter is presented in Fig.\ref{fig:fl} were
the $ud$-term  and the sum $ud + cs$ are shown separately  for different 
virtualities of the probe. For the $cs$-current $x=x_{Bj}(1+M^2_{cs}/Q^2)$
with $M^2_{cs}=4$ GeV$^2$.
Eqs.(\ref{eq:RHOS1},\ref{eq:RHOS2}) make it clear that  it is
 the  non-conservation of 
both   axial-vector and vector currents what
leads to the charm-strange dominance in  $F_L(x,Q^2)$ at small $x$.

\section{Summary and conclusions}

Summarizing, we considered the PCAC hypothesis in a specific domain of
 small Bjorken $x$,
where the relevant degrees of freedom are
the QCD color dipoles.
We reformulated Adler's theorem in the CD basis and analyzed
its efficiency  as a constraint requiring identical cross sections
for  scattering processes with pointlike and non-pointlike
probes.
This requirement, with certain reservations about
absorption/unitarity corrections, was found hard to  meet within 
the color dipole models successfully tested against HERA data. Corresponding
 non-perturbative
parameters including $m_q$ were adjusted to pave the way from the region 
$Q^2\simeq m_q^2$ to  high-$Q^2$ DIS. 
The analysis of diffractive vector mesons \cite{VecMesons} shows that the 
adjustment 
is not perfect.  
The discrepancy found in our paper can be understood  as the  non-perturbative
effect of 
increasing dipole size of the light-cone $W_L$ boson at $Q^2\to 0$.   
Adler's theorem  provides its quantitative  measure.
This observation     
 makes topical new experimental  tests
of Adler's theorem  in the diffraction region of $x\lsim 0.01$.
In view of the charm-strange dominance discussed above (which also
should be tested experimentally) 
 the 
$cs$-current contribution to the differential cross section
${d\sigma/dydQ^2}$ of  the reaction 
(\ref{eq:NU-N})
 has to  be isolated properly to separate the PCAC term.

\vspace{0.2cm} \noindent \underline{\bf Acknowledgments:}

Thanks are due  to N.N. Nikolaev for useful comments.
V.R.~Z. thanks  the Dipartimento di Fisica dell'Universit\`a
della Calabria and the Istituto Nazionale di Fisica
Nucleare - gruppo collegato di Cosenza for their warm
hospitality while a part of this work was done.
The work was supported in part by the Ministero Italiano
dell'Istruzione, dell'Universit\`a e della Ricerca
and  by
the DFG grant 436 RUS 17/82/06 and by the RFBR grant 06-02-16905-a.

\end{document}